\documentclass[journal]{IEEEtran}
\ifCLASSINFOpdf
\usepackage{amsmath}
\usepackage{graphicx}
\usepackage{float}
\usepackage{multirow}
\usepackage{array}
\usepackage{adjustbox}
\else
\fi
\begin{document}
%
\title{A Learned Pixel-by-Pixel Lossless Image Compression Method with 59K Parameters and Parallel Decoding}
%
%
%


\author{\IEEEauthorblockN{Sinem Gümüş, Fatih Kamisli}\\
\thanks{Both authors are with the Department of Electrical and Electronics Engineering at the Middle East Technical University, Ankara, Turkey. (emails: sinem.gumus@metu.edu.tr, kamisli@metu.edu.tr)}
}

\maketitle

\begin{abstract}
This paper considers lossless image compression and presents a learned compression system that can achieve state-of-the-art lossless compression performance but uses only 59K parameters, which is more than 30x less than other learned systems proposed recently in the literature. The explored system is based on a learned pixel-by-pixel lossless image compression method, where each pixel’s probability distribution parameters are obtained by processing the pixel’s causal neighborhood (i.e. previously encoded/decoded pixels) with a simple neural network comprising 59K parameters. This causality causes the decoder to operate sequentially, i.e. the neural network has to be evaluated for each pixel sequentially, which increases decoding time significantly with common GPU software and hardware. To reduce the decoding time, parallel decoding algorithms are proposed and implemented. The obtained lossless image compression system is compared to traditional and learned systems in the literature in terms of compression performance, encoding-decoding times and computational complexity.
\end{abstract}


%
\IEEEpeerreviewmaketitle

\section{Introduction} \label{sec:intro}

Image compression is the set of algorithms to reduce the number of bits to store or represent an image, with or without preserving the original image data perfectly. Lossless image compression preserves the original image data, i.e. guarantees obtaining the original image after decoding the compressed bitstream and is generally preferred in applications where data loss is not desired, such as medical imaging, professional photography, satellite imaging, etc. To achieve the compression, the statistical dependency of the pixel values along the spatial dimension and across the color channels are typically utilized.


Traditional lossless image compression methods are either based on prediction or integer transforms. In the prediction based methods, compression proceeds in a raster scan order and a pixel or block of pixels is predicted from previously compressed pixels or blocks available at the decoder and the prediction error is lossless coded with various methods \cite{pennebaker1992jpeg, weinberger2000loco, sneyers2016flif, webp, boutell1997png}. In the integer transform based methods, the image is transformed into a transform domain with an invertible integer to integer transform and the transform coefficients are lossless coded \cite{christopoulos2000jpeg2000}. In both methods, the spatial dependency of pixel values is exploited and error pixels or transform coefficients 
that can be more efficiently coded with simple entropy codes are obtained and coded to accomplish the compression.

Recently, learned lossless image compression methods have emerged. These methods use artificial neural networks to model and learn the probability distributions of pixels conditioned on previously coded pixels \cite{van2016pixel, oord2016conditional, salimans2017pixelcnn++} or side information transmitted from the encoder to the decoder \cite{mentzer2019practical, cao2020lossless, zhang2020lossless}. Methods that model and learn integer to integer transforms with artificial neural networks were also proposed \cite{hoogeboom2019integer}.

This paper explores a learned lossless image compression method which uses artificial neural networks to model and learn the probability distributions of each pixel conditioned on few previously coded left and upper neighbor pixels. In other words, a learned auto-regressive model is used to characterize the probability distribution of pixels. It is shown that a simple neural network with only 59K parameters can learn such an auto-regressive dependency very well and can achieve state-of-the-art lossless image compression performance that is better than other recent learned lossless image compression systems IDF\cite{hoogeboom2019integer}, L3C\cite{mentzer2019practical}, SReC\cite{cao2020lossless} and most models in MSPSM\cite{zhang2020lossless}. This result is surprising since many learned lossless compression approaches have number of parameters in the millions (IDF\cite{hoogeboom2019integer}: 84.3M, L3C\cite{mentzer2019practical}: 5.0M, SReC\cite{cao2020lossless}: 4.2M, MSPSM\cite{zhang2020lossless}: 1.9M) but can not achieve better lossless compression than this simple system. 

The successful compression performance with much less parameters is mainly due to two factors. First, in lossless image compression, few immediate neighbor pixels (i.e. a CNN with small receptive field) are sufficient for good performance, unlike lossy image compression or many computer vision tasks. Hence, our system uses convolution layers mostly with 1x1 kernels instead of larger kernels. Second, many systems in the literature operate in multiple scales (due to easy parallel computation possibility), which requires multiple neural networks, but our system does not.

The downside of our approach is that the auto-regressive model requires sequential processing of neighboring pixels that are related to each other by the auto-regressive model and a simple parallelization of the processing of all image pixels  (i.e. using convolutional neural networks) can not be used. A naive decoding implementation which processes all image pixels sequentially leads to impractically long decoding times with standard GPU hardware and software. Yet, by keeping the number of neighbor pixels related in the auto-regressive model small (i.e the neural network's receptive field), many pixel groups can be processed in parallel (i.e. decoded independently), and more reasonable decoding times can be achieved, which are still about an order of magnitude longer than fully parallelizable approaches such as L3C\cite{mentzer2019practical}, SReC\cite{cao2020lossless}, MSPSM\cite{zhang2020lossless}.

Contributions of this paper can be summarized as follows:
\begin{itemize}
    \item State of the art lossless image compression performance of 2.56 bpsp on the test set of Open Images dataset
    \item while using a neural network with only 59K parameters, which is 30x-1000x less than the parameters of systems in the literature.
\end{itemize}

The remainder of the paper is organized as follows. Section \ref{sec:relwrk} reviews related work including traditional lossless image compression methods and recent learned lossless image compression methods. Section \ref{sec:prop} presents the proposed approach and compression system. Section \ref{sec:expres} presents experimental results with the proposed system and also comparisons with other systems. Finally, Section \ref{sec:conc} provides a summary and our conclusions.

\section{Related Work}\label{sec:relwrk}
This section provides a review of related traditional lossless image compression methods and recent learned lossless image compression methods.

\subsection{Traditional Lossless Image Compression Methods} \label{ssec:trad}
Traditional lossless compression methods can be categorized as pixel based prediction methods and invertible transform based methods. In pixel based prediction methods, pixel values are predicted from their surrounding pixels (left and above) using predefined modes in the encoder. The difference signal is obtained by subtracting the prediction pixel from the original pixel value, and the obtained error pixel is lossless compressed using an entropy coder \cite{pennebaker1992jpeg, weinberger2000loco, sneyers2016flif, webp, boutell1997png}. Transformation based image coding methods are widely used in lossy compression. In lossless compression, the used transform must be invertible and satisfy integer-to-integer mapping in order to obtain the original image without loss at the decoder side \cite{christopoulos2000jpeg2000, liu2022improved, masmoudi2015improved}. 


JPEG \cite{pennebaker1992jpeg} provides a lossless operation mode which uses a pixel based predictive coding scheme for compression. The current pixel is predicted from the immediate neighbor pixels to the left, up, upper-left and upper-right. There are eight predefined prediction modes which are generated as different linear combination of neighbor pixels using only addition subtraction and bit-shift operations.

JPEG-LS \cite{weinberger2000loco} is based on the LOCO-I (Low Complexity Lossless Compression for Images) algorithm and is similar to lossless JPEG in terms of operation. It increases compression performance by using an enhanced decorrelation with adaptive prediction modes and context modelling. A nonlinear prediction algorithm, the median edge detector (MED) is used, which chooses between three estimation modes depending on edge information from the relationship between neighboring pixels. A context model, determined by quantized gradients of the neighboring pixels to capture high-order dependencies, is used with adaptive entropy coding.

FLIF (Free Lossless Image Format) \cite{sneyers2016flif} has one of the best compression performances among the traditional lossless compression systems. A reversible YCoCg color transformation \cite{malvar2003ycocg} is applied to decorrelate color channels. An entropy coding method called MANIAC (Meta-Adaptive Near-Zero Integer Arithmetic Coding) based on CABAC (Context-adaptive binary arithmetic coding) is used \cite{marpe2003context}. This method provides a powerful adaptive context model. Although FLIF offers strong compression performance, it has significantly longer run times than other traditional methods.

WebP \cite{webp}, developed especially for the web, offers lossy and lossless modes of operation. In the lossless mode, adaptive block based prediction with thirteen modes, a color transform and a variant of LZ77-Huffman coding is used for entropy coding. 

Another widely popular lossless image compression system is PNG \cite{boutell1997png}. Video compression standards H.264/AVC and HEVC also support lossless compression \cite{lee2006improved, zhou2012hevc}. Finally, JPEG2000 supports lossless compression with an integer wavelet transform based approach \cite{christopoulos2000jpeg2000}.

\subsection{Learning Based Lossless Image Compression Methods} \label{ssec:lear}
\subsubsection{Auto-regressive generative models} \label{sssec:argm} 
Auto-regressive generative models are a class of artificial neural networks whose main purpose is to learn the probability distribution of an image dataset and to generate/sample new images that appear to be in that set. For this purpose they perform auto-regressive generation of pixels. In order to use these models in lossless image compression, they need to be combined with an entropy coder.

PixelRNN \cite{van2016pixel}, PixelCNN \cite{oord2016conditional}, and MS-PixelCNN \cite{reed2017parallel} are well-known auto-regressive generative network architectures. The probability distribution function of the entire image is formulated as the joint probability distribution of its pixels, which is factored with the chain rule of probability into a product of conditional probabilities of each pixel conditioned on all previous pixels. PixelRNN\cite{van2016pixel} models large spatial dependencies in the image using the two-dimensional LSTM (long-short-term memory) network structure. PixelCNN \cite{oord2016conditional} offers a CNN (convolutional neural network) based architecture to speed up the training of the neural network and uses only certain neighboring pixels with the help of masked convolution structures. PixelCNN++ \cite{salimans2017pixelcnn++} proposes improvements on top of PixelCNN. In particular, it uses a discretized logistic mixture likelihood for the pixel probability distribution instead of the 256-way softmax function and a simpler method to learn the relation between RGB components of a pixel \cite{salimans2017pixelcnn++}.

Note that while the above discussed auto-regressive generative models are similar to the work in this paper in the sense that they also model/learn dependencies in images with auto-regressive neural networks, there are also significant differences. First, these models were not proposed particularly for lossless image compression but for image generation. Second, these models comprise very large neural networks with number of parameters in the millions (e.g. PixelCNN++: 53.7M) and their sampling times, which would correspond to decoder operations in lossless compression, are impractically long taking many minutes or hours with GPUs \cite{mentzer2019practical}. Such large neural networks and long sampling times render them highly impractical for lossless compression applications. This paper explores an auto-regressive lossless compression system with a much simpler neural network (59K parameters) and parallel decoding capability, which significantly reduce the computation complexity and long decoding times compared to the above discussed auto-regressive models while providing state-of-the-art compression performance.


\subsubsection{Parallel prior models} \label{sssec:ppm} 
Due to the sequential generation/sampling requirement and long decoding times with auto-regressive generative models, image probability models that allow the computation of probability parameters of each pixel in parallel were proposed \cite{mentzer2019practical, cao2020lossless, zhang2020lossless}. These approaches condition the probability of each pixel on a prior (i.e. side information) that is obtained and transmitted by the encoder to the decoder. This prior has typically half the resolution of the original image to be compressed and can be processed with convolutional neural networks (CNN), allowing simple parallelization. The decoder processes this prior with CNN and can produce the probability model parameters of each pixel in parallel. Note that the transmission of the prior will take some additional bits and thus to obtain an efficient compression framework, this prior based compression scheme is applied in a recursive manner on the priors, leading to hierarchical/multi-scale compression frameworks.

In L3C \cite{mentzer2019practical}, which is the first to propose such a framework, a 3-scale framework is used where the priors are obtained with CNNs at the encoder by processing the original image or the previous prior. The last prior is sent to the decoder with a simple entropy code. The decoder processes the received prior with CNN to obtain the probability model parameters of each element of the previous prior, and then decodes the previous prior. This procedure is repeated for the preceding prior and so on until the original image is decoded.

In super resolution based compression (SReC) \cite{cao2020lossless}, the compression framework is similar to L3C and the prominent difference is that the priors are obtained not with a CNN but simply by average pooling, i.e. averaging every 2x2 pixels. In MSPSM \cite{zhang2020lossless}, again a similar framework is used where the priors are obtained in a progressive manner simply by taking every even/odd rows and/or columns of the image or previous prior. 

While L3C provides practical encoding/decoding times with GPUs and compression performance higher than many traditional lossless image coders, the compression performance is below that of the advanced traditional coder FLIF \cite{sneyers2016flif}. SReC \cite{cao2020lossless} improves upon L3C in terms of compression performance and MSPSM \cite{zhang2020lossless} improves upon SReC.

While these methods provide practical encoding and decoding times with GPUs and have good compression performance they still are based on relatively large neural networks with high computational complexity. For example, L3C\cite{mentzer2019practical}, SReC\cite{cao2020lossless}, and MSPSM\cite{zhang2020lossless} have 5.0M, 4.2M and 1.9M parameters, respectively, while the system in this paper has only 59K parameters. 

\subsubsection{Integer Discrete flow based models}
Integer Discrete Flows (IDFs) are invertible deep learning based transformations for ordinal discrete data, such as images, similar to the traditional integer wavelet transforms \cite{christopoulos2000jpeg2000}. While such flow based models \cite{hoogeboom2019integer, vandenberg2021idf} provide good compression performance and allow simple parallelization through CNN, the used neural network architectures are quite large with number of parameters in the orders of many millions (e.g. 84.3M in \cite{hoogeboom2019integer}). Hence, these approaches seem inappropriate as practical lossless image compression solutions.

\section{Proposed Method} \label{sec:prop}
This paper explores a learned auto-regressive lossless image compression method which uses neural networks to model the conditional probability distribution of each pixel conditioned on previously encoded/decoded left and upper pixels. The remainder of this section is organized as follows. Section \ref{ssec:prob} discuses the used conditional probability model. Section \ref{ssec:netw} presents the used neural network architecture to learn the probability model parameters. Section \ref{ssec:paral} presents parallelization approaches to reduce decoding times.

\subsection{The Conditional Probability Model of Pixels} \label{ssec:prob}
\begin{figure}[th]
\centering
\includegraphics[scale=0.65]{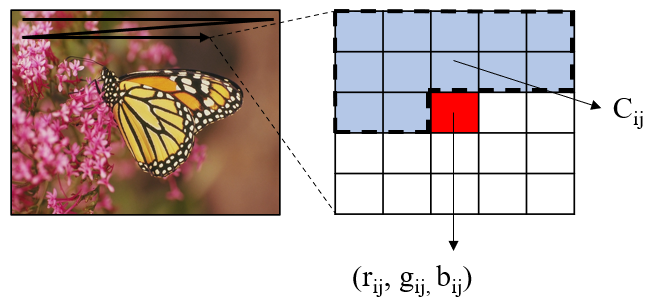}
\caption{Raster scan order for coding pixels. The pixel to be encoded next (red) and the neighboring previously encoded/decoded pixels $C_{i,j}$ (blue) that its probability model conditions on.}
\label{fig:pixelbased}
\end{figure}

The probability mass function (PMF) of a sub-pixel\footnote{Here, we follow the convention in the related previous research and use sub-pixel to denote each color component and pixel to denote all color components together} x $\in\{r,g,b\}$ is defined via discretizing a Gaussian Mixture Model (GMM) as follows:
\begin{equation}
p_x(x;~{{\pi,\mu,s}})=\sum _{i=1}^{K} \pi_{i}[F((x + 0.5 - \mu_{i})/s_{i})-F((x - 0.5 - \mu_{i})/s_{i})]
\label{eq:pr_sp}
\end{equation}
Here $F()$ is the cumulative distribution function (CDF) of the standard Gaussian distribution and $\mu_{i}$, $s_{i}$ and $\pi_i$ are the mean, standard deviation and the weight of the $i^{th}$ mixture/component of the GMM, respectively. A pixel at location (i,j) in the image consists of sub-pixels $r_{i,j}$ (red), $g_{i,j}$ (green) and $b_{i,j}$ (blue) sub-pixels. The conditional probability of a pixel is represented as the joint probability of these three sub-pixel probabilities as  
\begin{equation}
\begin{split}
P(r_{i,j} , g_{i,j} , b_{i,j} |C_{i,j}) = p_x(r_{i,j} ;~ \pi_{r}(C_{i,j}),\mu_{r}(C_{i,j}), s_{r}(C_{i,j})) \times \\ 
p_x(g_{i,j} ;~ \pi_{g}(C_{i,j}),\mu_{g}(C_{i,j} , r_{i,j}), s_{g}(C_{i,j})) \times \\ 
p_x(b_{i,j} ;~ \pi_{b}(C_{i,j}),\mu_{b} (C_{i,j} , r_{i,j}, g_{i,j}), s_{b}(C_{i,j}))
\end{split}
\label{eq:pr_rgb}
\end{equation}
where $C_{i,j}$ represents the previously encoded/decoded neighboring pixels that the probability model conditions on (pixels in blue region in Figure \ref{fig:pixelbased}). $\pi_{r}(C_{i,j}),\mu_{r}(C_{i,j})$ and $s_{r}(C_{i,j})$ represent the mean, standard deviation and the weight parameters for the sub-pixel $r_{i,j}$ (red) and are produced by the artificial neural network processing the $C_{i,j}$ pixels. As the $C_{i,j}$ pixels change, the parameters $\pi_{r}(C_{i,j}),\mu_{r}(C_{i,j})$ and $s_{r}(C_{i,j})$ change and thus the conditional probability of the sub-pixel $r_{i,j}$ changes.

The probabilities of green and blue sub-pixels $g_{i,j}$ and $b_{i,j}$ are determined by the mean, standard deviation and weight parameters shown in Equation (\ref{eq:pr_rgb}), which  are also obtained by the same artificial neural network processing the $C_{i,j}$ pixels. Notice however that the mean parameters $\mu_{g}(C_{i,j} , r_{i,j})$ and $\mu_{b} (C_{i,j} , r_{i,j}, g_{i,j})$ are functions of not only the $C_{i,j}$ pixels but the mean of the green sub-pixel is also a function of the previously encoded/decoded red sub-pixel $r_{i,j}$ and the mean of the blue sub-pixel is also a function of the previously encoded/decoded red and green sub-pixels $r_{i,j}$ and $g_{i,j}$. In other words, the probability of the green sub-pixel is conditioned on both the neighboring $C_{i,j}$ pixels and the red sub-pixel $r_{i,j}$ and the probability of the blue sub-pixel is conditioned on both the neighboring $C_{i,j}$ pixels and the red and green sub-pixels $r_{i,j}$ and $g_{i,j}$, which indicates that the joint probability model of the sub-pixels in Equation (\ref{eq:pr_rgb}) models the dependency of the sub-pixels through updating only the means in the GMM of Equation (\ref{eq:pr_sp}). The updates of the means are performed with the simple but efficient method of \cite{salimans2017pixelcnn++} as in Equations (\ref{eq:mnup_g}) and (\ref{eq:mnup_b}).

\begin{align}
\mu_{g}(C_{i,j}, r_{i,j}) = \mu_{g}(C_{i,j}) +\alpha(C_{i,j})r_{i,j}
\label{eq:mnup_g}
\end{align}
\begin{align}
\mu_{b}(C_{i,j} , r_{i,j} , g_{i,j}) = \mu_{b}(C_{i,j}) +\beta(C_{i,j})r_{i,j}+ \gamma(C_{i,j})g_{i,j}  
\label{eq:mnup_b}
\end{align}

Here, $\alpha(C_{i,j})$, $\beta(C_{i,j})$ and $\gamma(C_{i,j})$ are coefficients that multiply the red and green sub-pixels to update the means and are also produced by the same neural network processing the $C_{i,j}$ pixels. This simple method of updating the means is computationally efficient and is adequate to exploit the simple dependency of the sub-pixels. It is computationally efficient since one evaluation of the neural network processing the $C_{i,j}$ pixels produces all parameters necessary to calculate the joint conditional probability of the sub-pixels in Equation (\ref{eq:pr_rgb}). An alternative and straight forward method would be to have three neural networks, each taking as input the $C_{i,j}$ pixels and the previous sub-pixels but this would require three neural network evaluations and increase the neural network complexity and decoding time and is therefore not preferred.


\subsection{Neural Network Architecture and Loss Function} \label{ssec:netw}
\begin{figure}[th]
	\centering
	\includegraphics[scale=0.60]{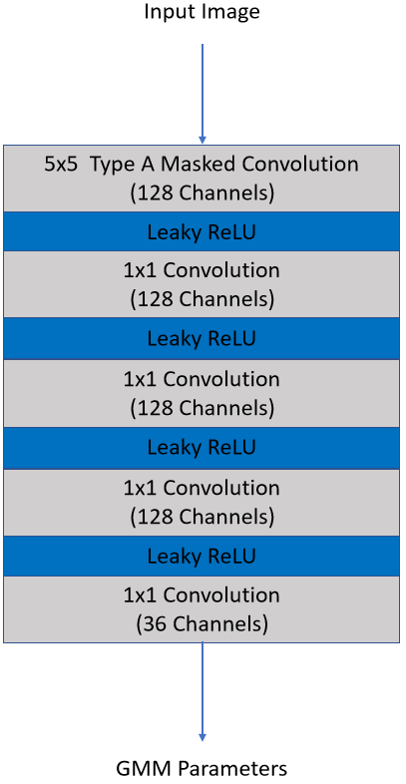}
	\caption{Neural network architecture of the proposed probability model.}
	\label{fig:network}
\end{figure}

The simple neural network used in this paper to obtain the conditional probability model parameters is shown in Figure \ref{fig:network}. It has 5 convolution layers and LeakyReLU activation function in between the convolution layers. The first convolution layer is a masked convolutional layer and has a kernel size of 5x5 while the remaining convolutional layers have kernel size of 1x1. Hence, the receptive field of the convolutional neural network is determined by the first layer's kernel size and mask, and comprise the $C_{i,j}$ pixels in Figure \ref{fig:pixelbased} that the conditional probability model of Equation \ref{eq:pr_rgb} conditions on. Note that one could have a larger receptive field $C_{i,j}$ by using larger kernel sizes in the convolutional layers. However, unlike in many computer vision tasks or lossy image compression, a larger receptive field is not necessary for state-of-the-art lossless image compression performance as we show in Section \ref{sec:expres} and the neural network in Figure \ref{fig:network} allows to keep the network parameter size and computational complexity low.


For the training of the proposed neural network, stochastic gradient descent is used and the loss function is simply the estimate of the entropy of the pixel values given below in Equation (\ref{eq:loss}), where $b$ indexes the images in a batch. More details of the training procedure are given in Section \ref{sec:expres}.
\begin{equation} 
 L = -{\frac  {1}{N}} \sum _{{b,i,j}}\log _{2} P(r^{(b)}_{i,j},g^{(b)}_{i,j},b^{(b)}_{i,j}| C_{i,j})
 \label{eq:loss}
\end{equation}

\subsection{Parallelization of Decoding Operations} \label{ssec:paral}
During training or encoding, since all image pixels are available, the entire image can be fed to the CNN in Figure \ref{fig:network} and all pixels can be processed in parallel to obtain their probability distribution parameters. The masked convolution ensures that the auto-regressive dependency on only the left and upper $C_{i,j}$ pixels is preserved. During decoding, however, sequential processing is required. To decode one pixel from the bitstream, that pixel's probability distribution parameters need to be obtained by processing the previously decoded $C_{i,j}$ pixels with the neural network. Once this pixel is decoded from the bitstream, it is used in the $C_{i,j}$ pixels of the next pixel and the process for decoding the next pixel is performed similarly.

During decoding, the evaluation of the neural network to obtain the probability distribution parameters of one pixel can be performed in two similar ways. One way is to form a 5x5 patch of pixels containing the previously decoded $C_{i,j}$ pixels as shown in Figure \ref{fig:pixelbased} and feed it to the convolutional neural network in Figure \ref{fig:network}. An alternative way is to form a full-connected neural network using the weights of the convolutional layers in Figure \ref{fig:network} and input the previously decoded $C_{i,j}$ pixels to this network. Either method will give the same result.

\begin{figure}[h]
\centering
\includegraphics[scale=0.60]{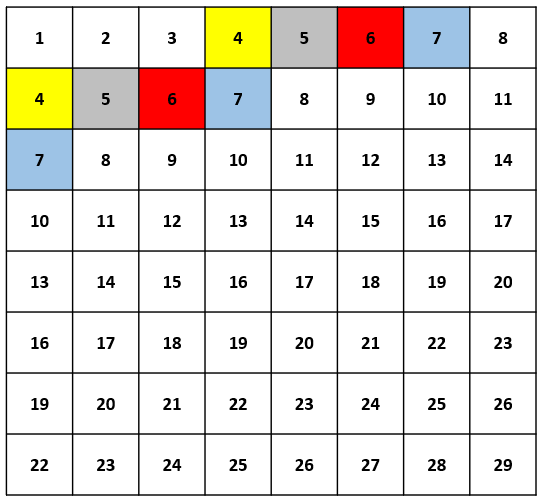}
\caption{Decoding order of pixels with the wavefront parallel decoding method for kernel size 5x5.}
\label{fig:parallel_dec1}
\end{figure}

Notice that the decoding process does not need to be sequential for all pixels in the image. Many pixel groups can be identified for which the pixels can be processed in parallel with the neural network. All pixels for which the $C_{i,j}$ have been decoded can be processed in parallel. Since we kept the receptive field of the neural network (i.e. $C_{i,j}$) small, a great deal of parallelization is possible. The numbers in Figure \ref {fig:parallel_dec1} indicate the order in which the pixels are decoded by the decoder and the pixels which have the same color can be processed in parallel by the neural network. As an example, consider an image with D x D resolution. For a fully sequential decoder, each pixel is processed sequentially and it takes $D^{2}$ forward passes to decode the entire image. The parallelization method in Figure \ref{fig:parallel_dec1}, which has been used in many applications and is also called parallel wavefront decoding in the HEVC video compression standard \cite{chi2012parallel, zhang2022parallel}, reduces number of passes from $D^{2}$ to:


\begin{equation} \label{eq:parallel_1}
 T = D+(D-1)(h+1)
\end{equation}
where $h$ is an integer related to the kernel size $K = 2h + 1$.

\begin{figure}[h]
\centering
\includegraphics[scale=0.60]{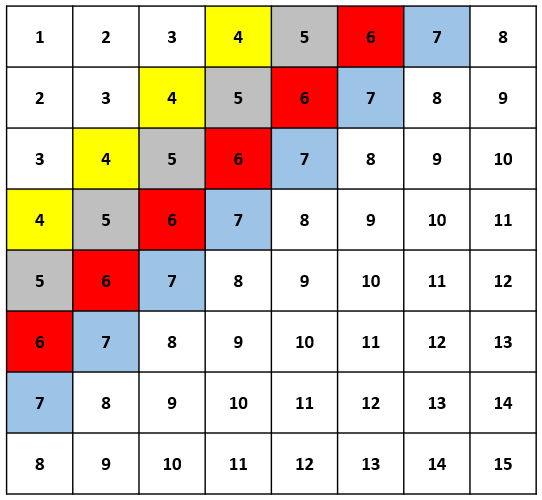}
\caption{Decoding order of pixels with the diagonal parallel decoding method.}
\label{fig:parallel_dec_diag}
\end{figure}

The wavefront parallel decoding method can be further improved by compromising the use of some of the neighboring pixels. The diagonal parallel decoding method shown in Figure \ref{fig:parallel_dec_diag} can decode all pixels along a diagonal in parallel. The number of passes reduce to
\begin{equation} \label{eq:parallel_2}
 T = 2D-1.
\end{equation}

\begin{figure}[h]
\centering
\includegraphics[scale=0.60]{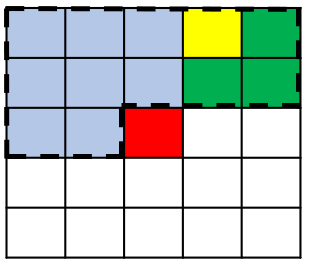}
\caption{Unavailable neighbor pixels (green pixels) for the diagonal parallel decoding method.}
\label{fig:parallel_dec_unavaliable}
\end{figure}

The diagonal parallel decoding method increasing parallelization but compromises compression performance since some neighbor pixels cannot be utilized in this parallelization method. As shown in Figure \ref{fig:parallel_dec_unavaliable}, three pixels in the upper diagonal (shown in green) can not be used as they are not yet decoded and their values are simply copied from the yellow pixel in the figure.


Finally, note that if a parallel decoding method is to be used, the encoder needs to be aware and its entropy coder needs to encode the pixels in the order the parallel decoding method will decode.

\section{Experimental Results} \label{sec:expres}
This section provides experimental results including compression performance results, encoding/decoding times, computational complexity estimate and an ablation study.

\subsection{Experimental Setting} \label{ssec:expset}
The simple neural network in our system is trained with sub-images obtained from the Open Images training dataset prepared by the authors of L3C \cite{mentzer2019practical}. The Adam optimizer \cite{kingma2014adam} is used for optimization with a batch size of 64. The learning rate is initialized as $10^{-4}$ and updated by multiplying with $0.99$ after every five training epoch. The training takes about 300 epochs and $16.5M$ iterations. PyTorch \cite{paszke2019pytorch} framework is used for both training and testing. 

Our system is tested with the test dataset prepared again by the authors of L3C \cite{mentzer2019practical}, which includes 500 images. The arithmetic coder in \cite{numpyAc}, which runs on the CPU, is used in our system for encoding to and decoding from a bitstream during the tests. Note that all neural network operations are performed on the GPU and the arithmetic coding operations are performed on the CPU in our tests. Test results are obtained with a computer which has an NVIDIA GeForce RTX 2080 GPU and Intel i7-9700 CPU. Our codes are available on Github at the link in \cite{sgumus}. 

\subsection{Compression Results} \label{ssec:comres}
The compression performance results for the test dataset are given in Table \ref{tb:compres} in terms of bpsp (bits per sub-pixel). The learned pixel-by-pixel lossless image compression (LPPLIC) system explored in this paper obtains an average 2.56 bpsp compression performance and outperforms all engineered codecs, including FLIF, by a significant margin. It also provides better compression performance than all learned lossless compression methods, except MSPSM(extra). However this system requires 9.9M parameters while our system LPPLIC requires only 59K parameters, which is 168x less. 

In summary, the explored LPPLIC system provides very competitive compression performance with a much simpler neural network comprising significantly less parameters, in particular 32x to 1430x less parameters than other learned systems in Table \ref{tb:compres}.

\begin{table}[ht]
\centering
\caption{\textsc{Compression Results}}
\label{tb:compres}
\begin{tabular}{ c|c|c|r } 
\hline
 & Method & bpsp & Number of\\
 &        &      &  Parameters \\
\hline
\multirow{4}{5em}{Traditional Methods} 
& PNG & 4.01 & - \\ 
& JPEG2000 & 3.06 & - \\
& WebP & 3.05 & -  \\
& FLIF & 2.87 & - \\
\hline
\multirow{7}{5em}{Learning Based Methods} 
& L3C \cite{mentzer2019practical}       & 2.99 & 5.0 M \\ 
& IDF \cite{hoogeboom2019integer}       & 2.76 & 84.3 M \\
& SREC \cite{cao2020lossless}           & 2.70 & 4.2 M \\
& MSPSM(big) \cite{zhang2020lossless}   & 2.63 & 1.9 M \\
& MSPSM(extra) \cite{zhang2020lossless} & 2.49 & 9.9 M \\
& LPPLIC (Our method)                      & 2.56 & 63.9 $\mathbf{K}$ \\
\hline
\end{tabular}
\end{table}

\subsection{Encoding and Decoding Times} \label{ssec:times}
During encoding, since all image pixels are available at the encoder, all image pixels can be processed in parallel with a single pass of the entire image through the convolutional neural network to obtain their probability distribution parameters. The masked convolution ensures that the auto-regressive dependency on only the left and upper pixels is preserved. From the obtained probability distribution parameters, cumulative distribution function (CDF) values are calculated for all pixels (in parallel) and are handed over to the entropy coder together with the pixel values to be coded. Note that even if a parallel decoding method is to be used, the encoder can still do all operations in the same way; only the order of coding the pixels need to be changed by the entropy coder to match the order the decoder will need to decode. Hence, practical encoding times can be achieved with the explored LPPLIC system as shown in Table \ref{tb:times}. 

\newcolumntype{C}[1]{>{\centering\arraybackslash}p{#1}}

\begin{table}[ht]
\centering
\caption{\textsc{Encoding/Decoding Times (sec) of our method LPPLIC with various decoding methods vs L3C and SReC}}
\label{tb:times}
\begin{adjustbox}{width=\columnwidth,center}
\begin{tabular}{l|c|c|c|c|c}
\hline
 Resolution & \multicolumn{1}{c|}{Sequential} & \multicolumn{1}{c|}{Wavefront} & \multicolumn{1}{c|}{Diagonal}   & \multicolumn{1}{c|}{L3C\cite{mentzer2019practical}} & SReC\cite{cao2020lossless}\\
            & \multicolumn{1}{c|}{decoding}   & \multicolumn{1}{c|}{parallel}  & \multicolumn{1}{c|}{parallel}   & \multicolumn{1}{c|}{}                               & \multicolumn{1}{c}{}       \\
            & \multicolumn{1}{c|}{}           & \multicolumn{1}{c|}{decoding}  & \multicolumn{1}{c|}{decoding}   & \multicolumn{1}{c|}{}                               & \multicolumn{1}{c}{}   \\ \hline
 64x64   & \multicolumn{1}{c|}{0.08 / 11.57}  & \multicolumn{1}{c|}{0.89 / 1.39}  & \multicolumn{1}{c|}{0.91 / 1.16}  & \multicolumn{1}{c|}{0.14 / 0.11} & 0.13 / 0.14\\ \hline
 128x128 & \multicolumn{1}{c|}{0.24 / 45.94}  & \multicolumn{1}{c|}{1.05 / 3.04}  & \multicolumn{1}{c|}{1.04 / 2.82}  & \multicolumn{1}{c|}{0.16 / 0.13} & 0.15 / 0.19\\ \hline
 256x256 & \multicolumn{1}{c|}{0.53 / 181.13} & \multicolumn{1}{c|}{1.51 / 7.93}  & \multicolumn{1}{c|}{1.52 / 7.72}  & \multicolumn{1}{c|}{0.24 / 0.20} & 0.20 / 0.32\\ \hline
 512x512 & \multicolumn{1}{c|}{1.64 / 711.03} & \multicolumn{1}{c|}{3.10 / 26.26} & \multicolumn{1}{c|}{3.08 / 27.22} & \multicolumn{1}{c|}{0.49 / 0.54} & 0.54 / 0.62\\ \hline
\end{tabular}
\end{adjustbox}
\end{table}

Note that a time overhead is observed when the image is compressed according to the parallel decoding methods as shown in Table \ref{tb:times}. Additional time losses occur due to pre-processing steps being applied at the encoder to correctly decode the image on the decoder side. 

During decoding, processing of all pixels can not be done in parallel as discussed in Section \ref{ssec:paral}. Either all pixels are decoded sequentially or one of the two parallel decoding methods discussed in Section \ref{ssec:paral} can be used. For all decoding methods, the decoding times are given in Table \ref{tb:times}. 
In the sequential decoding method, pixels are decoded one by one, each requiring one pass through the neural network and the decoding time becomes quite long, such as 711 seconds for an 512x512 image. With the wavefront or diagonal parallel decoding methods discussed in Section \ref{ssec:paral}, many groups of pixels can be processed in parallel (which we achieve by simply combining these pixels and their neighbors in the batch dimension of tensors) and fewer passes through the neural network are required leading to significant reduction in the decoding times. For example, for an 512x512 image, the decoding times reduce to 26 and 27 seconds with the wavefront and diagonal parallel decoding methods, respectively.

Table \ref{tb:times} includes encoding and decoding times also for L3C \cite{mentzer2019practical} and SReC \cite{cao2020lossless} (obtained with our computer using their shared codes) which are much smaller than those of LPPLIC. This is due to these systems' computations being easily parallelizable on GPU and not requiring evaluations with neural networks for each pixel to be encode/decoded, as discussed in Section \ref{sssec:ppm}. Note that Table \ref{tb:times} includes encoding and decoding times for various resolution images in order to also present how the times change with varying resolution. The numbers in the table were obtained by averaging the encoding/decoding times over ten images.

In summary, the explored LPPLIC method provides (with our non-optimized simple implementation) encoding times that are about 6x times longer than the fully parallelized methods such as L3C and SReC and decoding times that are about 42-48x times longer. In applications where state-of-the art compression performance and orders of magnitude less parameters and lower neural network complexity are more important than short encoding/decoding times, the explored LPPLIC approach can be preferred.

\subsection{Computational Complexity}
Since the number of parameters of the neural network in the explored LPPLIC method is much smaller than other similar learned systems, such as L3C \cite{mentzer2019practical} and SReC \cite{cao2020lossless}, it is also expected that the computational complexity of compressing/decompressing images with LPPLIC is also much smaller. To investigate this, the FLOP counter tool in \cite{flops} is used for LPPLIC and L3C and the computational complexity estimates for encoding/decoding a 512x512 image are given in Table \ref{tb:flop}. As expected the computational complexity of LPPLIC is much smaller. Note that despite the longer decoding times of LPPLIC, it requires significantly less computations.

\begin{table}[h]
\centering
\caption{MAC (Multiply-Accumulate) Operations Estimate\\ for a 512x512 image}
{\begin{tabular}{l|c|c}
 \hline
             & L3C \cite{mentzer2019practical} & LPPLIC \\[0.5ex]     \hline
    Encoding & 179.7 GMac                      & 16.9 GMac  \\     \hline
    Decoding & 112.3  GMac                      & 16.9 GMac  \\     \hline
\end{tabular}}
\label{tb:flop}
\end{table}

\subsection{Ablation Study} \label{ssec:abl}
An ablation study is performed to present the effects of different hyper-parameters of the LPPLIC system and associated neural network on compression performance and number of parameters. The hyper-parameters of the LPPLIC system that was used to present the compression and encoding/decoding time results in the previous two sub-sections, which were given in Figure \ref{fig:network} and are repeated in Table \ref{table:reference_params}, are analyzed 
for their effect on compression performance and parameter size. In the analyses below, only the discussed hyper-parameter is modified while the remaining ones are kept as in the Table \ref{table:reference_params}.
\begin{table}[h]
\centering
\caption{Hyper-parameters in the model}
{\begin{tabular}{c|c}
 \hline
    & Hyper-parameter \\[0.5ex] 
    \hline
    Probability distribution & Gaussian Mixture Model  \\ \hline
    \# mixtures & 3  \\ \hline
    \# filters in CNN layers & 128 \\ \hline
    \# CNN layers & 5 \\ \hline
\end{tabular}}
\label{table:reference_params}
\end{table}

\subsubsection{Probability distribution model} \label{sssec:prob}
A common alternative to the used Gaussian Mixture Model is the Logistic Mixture Model where the logistic cumulative distribution function is given below and $\mu$ and $s$ are the mean and scale parameter of the logistic distribution. 
\begin{equation}
F_x(x;~{{\mu,s}})= \dfrac{1}{1+e^{-\frac{x-\mu}{s}}}
\label{eq:logis}
\end{equation}
The Logistic and Gaussian probability distributions are similar in shape, and no major difference is expected in terms of compression results. As a result of the experiment, it was observed that using the Gaussian distribution resulted in a slightly better compression results as shown in Table \ref{table:gauss_vs_logistic}.
\begin{table}[h]
\centering
\caption{The Effect of different probability distributions}
{\begin{tabular}{c|c|c}
 \hline
     & Gaussian & Logistic \\[0.5ex] 
    \hline
    bpsp & 2.563 & 2.607 \\
    \hline
\end{tabular}}
\label{table:gauss_vs_logistic}
\end{table}

\subsubsection{Number of mixtures} \label{sssec:nummix}
The number of mixtures used in the Gaussian Mixture Model is an important parameter of the probability distribution. It can be seen from the results in  Table \ref{table:numer_of_mix} that increasing the number of mixtures beyond 3 does not significantly improve the compression performance. The number of trainable parameters is also given in the table. The number of mixtures increases the number of parameters of only the final convolutional layer and does not significantly change the number of parameters of the overall neural network. Based on these results, a mixture size of $K=3$ was used in the experimental results. 
\begin{table}[h]
\centering
\caption{The Effect of number of mixtures in GMM}
{\begin{tabular}{c|c|c|c}
 \hline
    Number of mixtures & K=3 & K=5 & K=7 \\[0.5ex]     \hline
    bpsp & 2.563 & 2.548 & 2.545 \\     \hline
    \# parameters  & 58917 & 62013 & 65109 \\     \hline
\end{tabular}}
\label{table:numer_of_mix}
\end{table}

\subsubsection{Number of filters in CNN layers} \label{sssec:numchs}
In order to analyze the number of filters in CNN layers on compression performance, experiments with different number of filters were performed. Based on the obtained results in Table \ref{table:channel_size}, increasing the number of filters does improve compression performance with diminishing gains but the number of parameters of the neural network increases more significantly (since the number of parameters increases with the square of the number of filters as a rough estimate.) Based on these results, C=128 filters were used in the experimental results. Note that if the number of parameters or computational complexity of the neural network is highly important for an application, smaller number of filters such as C=64 can also be used at the expense of some reduction in compression performance. 
\begin{table}[h]
\centering
\caption{The Effect of number of filters in CNN layers}
{\begin{tabular}{c|c|c|c}
 \hline
    Number of filters & C=64 & C=128 & C=192 \\[0.5ex]     \hline
    bpsp & 2.619 & 2.563 & 2.529 \\     \hline
    \# parameters & 17189 & 58917 & 125221 \\     \hline
\end{tabular}}
\label{table:channel_size}
\end{table}

\subsubsection{Number of layers in CNN} \label{sssec:numlys}
In order to analyze the number of layers in the CNN on compression performance, experiments with different number of layers were performed. Based on the obtained results in Table \ref{table:layers}, $L=5$ layers were chosen for a reasonable balance between compression efficiency and number of parameters.
\begin{table}[h]
\centering
\caption{The Effect of number of layers in CNN}
{\begin{tabular}{c|c|c|c}
 \hline
    Number of CNN layers & L=4 & L=5 & L=6 \\[0.5ex]     \hline
    bpsp & 2.572 & 2.563 & 2.543 \\     \hline
    \# parameters & 42405 & 58917 & 75429 \\     \hline
\end{tabular}}
\label{table:layers}
\end{table}

\section{Conclusions} \label{sec:conc}
This paper presented a learned lossless image compression method with pixel-by-pixel processing. The probability distribution of each pixel was obtained by processing a small causal neighborhood of it (i.e. few previously encoded/decoded left and upper pixels) with a simple neural network, which could then be used by an arithmetic encoder/decoder to encode/decode the pixel. It was shown that a simple neural network comprising only 59K parameters was sufficient to obtain state of the art compression performance in this system. Other learned lossless compression systems in the literature achieved similar or inferior compression performance with number of parameters that are at least 30x-1000x more. While the pixel-by-pixel processing causes the decoder to operate sequentially, i.e. the neural network has to be evaluated for each pixel sequentially, which increases decoding time significantly, parallelization methods were proposed -- thanks to the small receptive field of the used neural network -- and the decoding times were reduced to more reasonable levels. 
Overall, compared to the literature, better compression performance with a significantly smaller neural network and computational complexity was achievable at the expense of increased decoding times due to the decreased parallel computational capabilities of the pixel by pixel processing method.

\ifCLASSOPTIONcaptionsoff
  \newpage
\fi



%
\bibliographystyle{IEEEtran}
\bibliography{thesis.bib}{}


%







\end{document}